\RequirePackage{fix-cm}
\documentclass[smallextended]{svjour3}       
\smartqed  
\usepackage{graphicx}
\usepackage[numbers]{natbib}
\usepackage{amsmath}
\usepackage[center]{caption}
\usepackage[version=4]{mhchem}
\usepackage{siunitx}
\usepackage{graphicx}
\usepackage{caption}
\usepackage{hyperref}
\usepackage{amsmath}
\usepackage[justification=centering]{caption}
\begin{document}

\title{ComStreamClust: a communicative multi-agent approach to text clustering in streaming data}

\subtitle{}


\author{Ali Najafi  \and
        Araz Gholipour-Shilabin \and
        Rahim Dehkharghani \and
        Ali Mohammadpur-Fard \and
        Meysam Asgari-Chenaghlu
}


              
 \institute{Ali Najafi \at
               Department of Computer Engineering, University of Tabriz \\
               \email{najafi1998ali@gmail.com}           
           \and
           Araz Gholipour-Shilabin  \at
               Department of Computer Engineering, University of Tabriz\\
               \email{shilabin78@gmail.com} 
               \and
          Rahim Dehkharghani  \at
               Department of Computer Engineering, University of Bonab\\
               \email{rdehkharghani@ubonab.ac.ir} 
               \and
          Ali Mohammadpur-Fard  \at
               Department of Computer Engineering, Sharif University of Technology\\
               \email{ali.mpfard@gmail.com} 
               \and
          Meysam Asgari-Chenaghlu  \at
               Department of Computer Engineering, University of Tabriz\\
               \email{m.asgari@tabrizu.ac.ir} 
 }

\date{Received: date / Accepted: date}

\maketitle

\begin{abstract}
Topic detection is the task of determining and tracking hot topics in social media. Twitter is arguably the most popular platform for people to share their ideas with others about different issues. One such prevalent issue is the COVID-19 pandemic. Detecting and tracking topics on these kinds of issues would help governments and healthcare companies deal with this phenomenon. In this paper, we propose a novel, multi-agent, communicative clustering approach, so-called ComStreamClust for clustering sub-topics inside a broader topic, e.g., COVID-19. The proposed approach is parallelizable, and can simultaneously handle several data-point. The LaBSE  sentence embedding is used to measure the semantic similarity between two tweets. ComStreamClust has been evaluated on two datasets: the COVID-19 and the FA CUP. The results obtained from ComStreamClust approve the effectiveness of the proposed approach when compared to existing methods.
\end{abstract}

\section{Introduction}
\label{intro}
Social media, which has achieved growing popularity in recent decades, provide the opportunity for people to share their ideas with an enormous number of users around the world. 
As a micro-blogging platform, Twitter allows its users to write short text messages regarding various issues ranging from politics, economy, and healthcare to routine tasks of people's daily lives. One such issue, the COVID-19 pandemic, has had a profound impact on people's social lives since the beginning of 2020.

Determining and tracking health issues such as COVID-19 on Twitter would help governments and healthcare companies better handle the impact of those diseases on societies. Concretely, assembling tweets on this topic and analyzing them may result in invaluable information for those companies. From the healthcare perspective, crawling tweets related to COVID-19 as a pandemic issue might help in finding a remedy for it.
As manual processing of such information is prohibitively expensive, automatic or semi-automatic methods are thus needed; however, assembling and distilling such data is a challenging task.

Previous works have tackled this problem by streaming and grouping tweets into various categories by using supervised \cite{dehkharghani2014sentimental} or unsupervised \cite{cataldi2010emerging} methods. Unsupervised methods, however, could gain greater popularity. These methods collect streaming tweets in a time interval and assign them to clusters based on their topics. 

Clustering has been already used for topic detection in the literature. In stream data clustering, a two-phase task is accomplished. In the first phase, data are captured from a data stream; and in the second phase, clusters are created and (in this paper) re-organized to constitute denser clusters. The ultimate goal is to increase the intra-cluster similarities as well as to decrease the inter-cluster similarities.

To tackle the aforementioned problem, we propose a novel, communicative, multi-agent, parallelizable text clustering approach for tweet clustering, experimented on the COVID-19 and the FA CUP datasets, which is described with greater details in Section \ref{proposedMethod}. 
The key aspect of this work is its multi-agent and communicative structure.
The difference between this work and the existing ones is in the second phase (as mentioned above). In the communication step of the proposed approach, existing clusters may export data to, and/or import data from other clusters. At the same time, the proposed approach is also capable of distinguishing outlier data and excluding them from their current clusters. All these tasks can be (and have been) accomplished in a parallel setting. The contributions of the proposed approach can be summarized as follows. 
\begin{itemize}
\item ComStreamClust updates clusters by detecting outliers and distributing them among other clusters, in a streaming, parallel, and multi-agent setting. This setting is being used for the first time in the literature on topic detection problems.
\item The proposed approach could achieve promising results when applied to the FA CUP dataset. We applied our approach also to this dataset for the sake of fair comparison. Obtained results were as good as or superior to the existing approaches such as LDA, SFPM, and BNgram.
\item ComStreamClust benefits from a state-of-the-art sentence embedding model, the LaBSE, for measuring the semantic similarity between tweets.
\item A comprehensive experimental evaluation of the proposed approach on two datasets with different parameter values, such as the number of topics per time-slot, and the number of keywords per topic have been conducted.
\end{itemize}

\section{Related Work}
\label{RW}
Ibrahim et. al. \cite{ibrahim2018tools} divide the topic detection techniques into five groups: clustering, frequent pattern mining, Exemplar-based, matrix factorization, and probabilistic models. The current research falls into the clustering-based models. Stream clustering is a type of clustering, in which, data are continuously fed to a clustering system. Given this sequence of data, the goal is to group them in clusters, the elements of which are similar to each other but different from the elements of other clusters. In \cite{petkos2014two}, the authors propose a two-level clustering method based on document-pivot algorithm to detect topics in Twitter streaming data.  

Some previous work relies on word frequency for topic detection and topic categorization on Twitter data. Using the ``aging'' theory for modeling the term life-cycle has experimented in \cite{cataldi2010emerging}. In this work, Cataldi et.al., define a term as emerging, if it was rare in the past but frequent in a specified time-interval. The authors benefit from emerging terms to detect emergent topics. In \cite{dehkharghani2014sentimental} and \cite{dehkharghani2013automatically}, the authors propose approaches that detect topics in Twitter, based on their word frequency, and categorize tweets into sentiment classes. 

Unlike traditional clustering approaches that rely on a fixed set of input data, stream clustering assumes that input data are in a stream with an unknown number of usually unlabelled data. Along with the fast growth of social media such as Twitter, stream text clustering has gained growing popularity in recent decades. Several researchers have tackled this problem with different approaches. The proposed method in \cite{carnein2017stream} incrementally builds micro-clusters, 
which are later re-clustered to assemble final clusters. The idea of micro-clusters was earlier used in \cite{hahsler2016clustering}, in which, the
first micro-cluster-based online clustering algorithm, so-called DBSCAN was introduced. 
Hasler and Bolanos in this work, took into account the density of area between the micro-clusters, for the first time in the literature.

In another perspective, Fang et. al., \cite{fang2014detecting} categorize topic detection methods in Twitter into two main groups: traditional and new topic detection methods. In the traditional side, some research works \cite{guo2013lda} use an extension of LDA \cite{blei2003latent} for solving the topic detection problems. Some others tackle this problem by constructing a term co-occurrence network of keywords \cite{zhou2011hot} and single-pass clustering along with a new threshold method \cite{papka1998line}. 

Although traditional methods work well for long texts, they do not portray high performance on short texts such as tweets. Therefore, in new topic detection methods, traditional approaches have been extended to deal with new data types. In \cite{fang2014detecting}, the authors propose a new topic detection framework based on multi-view clustering. The Gradient Boosted Decision Trees is another method for detecting controversial events from Twitter which has been used in \cite{popescu2010detecting}.
Computational cost is one of the major challenges in the real-time topic- or event-detection on Twitter. Hasan et. al. \cite{hasan2019real} deal with this issue by proposing an event-detection system called TwitterNews+ which utilizes inverted indices and an incremental clustering approach, with a low computational cost. Asgari et. al \cite{asgari2020COVID} propose a model based on the universal sentence encoder \cite{cer2018universal} and transformers \cite{vaswani2017attention} to detect main topics on Twitter regarding the COVID-19 pandemic. 

Early detection of bursty topics is one of the most challenging problems in this era. TopicSkech \cite{xie2016topicsketch} deals with this problem on Twitter. Similarly, PoliTwi \cite{rill2014politwi}, also has been proposed for early detection of emerging political events. There are several other approaches to topic detection on the text which use different methods such as Formal Concept Analysis \cite{cigarran2016step}, clustering based on n-grams, and named entity boosting \cite{tembhurnikar2015topic}, and combination of singular value decomposition and K-means clustering methods \cite{nur2015combination}. More information and related work on topic detection on Twitter can be found in \cite{atefeh2015survey,mottaghinia2020review}.

Despite a good deal of research work proposed for topic detection on Twitter, none of them is communicative, multi-agent, and parallelizable at the same time. The proposed approach can improve its clusters by providing the communication ability for its clusters.

\section{Proposed approach}
\label{proposedMethod}
Twitter streaming data is a sequence of data, in which, data-points appear during the time. The problem tackled in this paper can be formally defined as follows:
Each data-point is assumed as a quadruple $(id, t, ts, s)$, such that \emph{id} is a unique value as the identification number; \emph{t} is the text with at most 280 characters; \emph{ts} is the timestamp of the tweet including its arrival date and time; \emph{s} is the subject of the tweet which is not known in advance. Once the subject \emph{s} is determined, the tweet can be assigned to one of the existing clusters. Having a set of topic clusters, the task is to assign a newly arrived tweet to one of the clusters. After this assignment, the attribute \emph{s} of the tweet will be initialized, which can be updated later.

\begin{figure*}[t]
\centering
\caption{The proposed approach as a flowchart.}
\includegraphics[width=12cm]{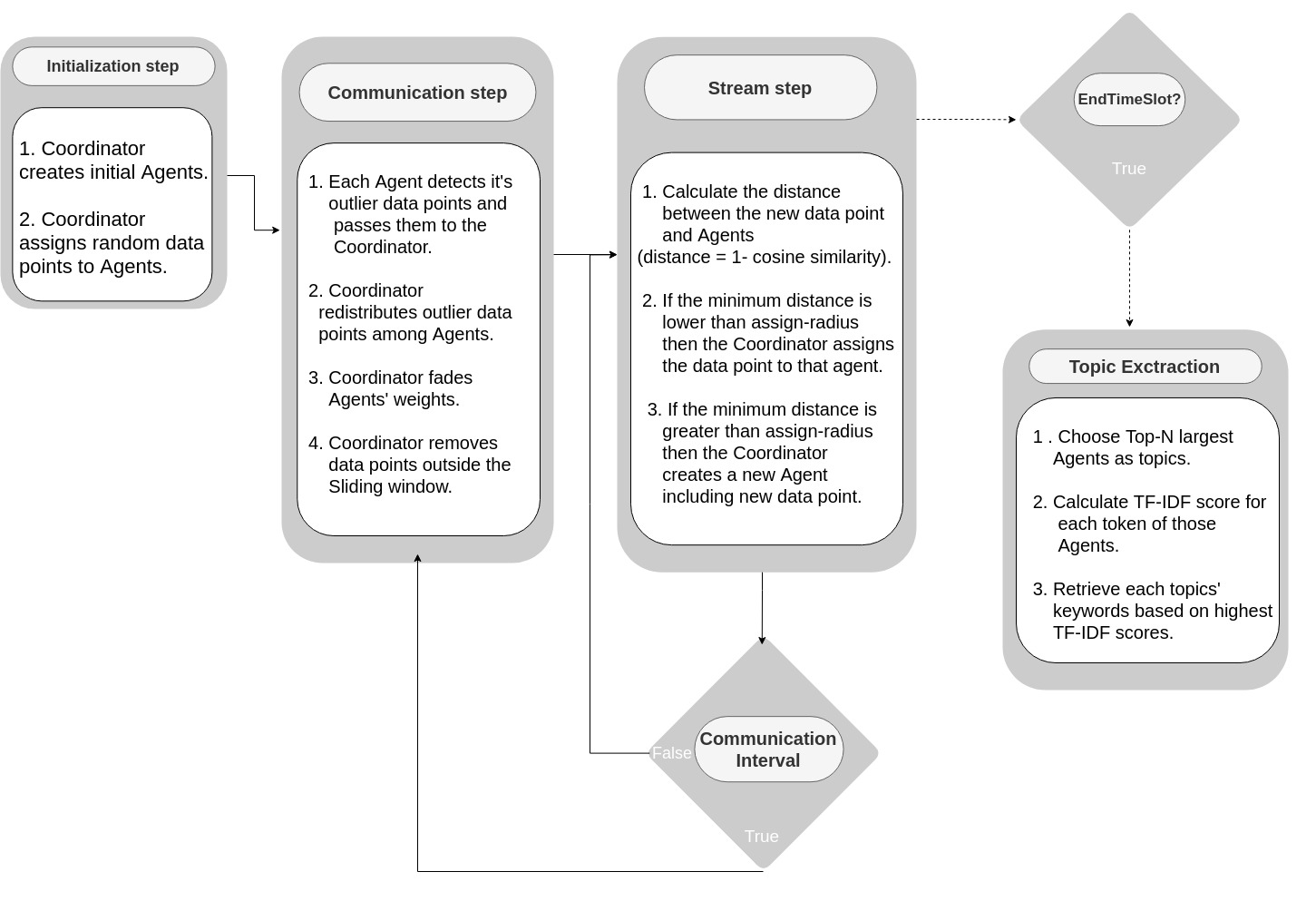}
\captionsetup{justification=centering}
\label{flowchart}
\end{figure*}

ComStreamClust consists of three main steps: (1) Data streaming, in which, data points, i.e. tweets, are fetched from Twitter streaming data, (2) Data assignment, where the newly arrived tweet is assigned to an existing or a new cluster, and (3) Data exchange, which is a communicative step to exchange data among clusters to build denser clusters.
The initialization phase is not assumed as the main step because it is accomplished once at the beginning of the whole process. Different steps of the proposed approach are described in greater detail in their respective subsections. The proposed approach is illustrated in figure \ref{flowchart} as a flowchart. The reference implementation of the proposed approach is also released under the MIT license\footnote{The Python and Elixir imlemantation of the proposed approach is publicly available at \url{https://github.com/AliNajafi1998/ComStream}}.

\subsection{Initialization}
\label{Initialization}
The initialization phase receives and handles the first \emph{k} tweets; we set \emph{k} to \emph{10}. The first \emph{k} tweets are randomly assigned to one of the initial agents. This phase will be finished when all initial agents are filled. The initial number of agents, identified by the \emph{init-agents} parameter, and the initial capacity of these agents indicated by \emph{init-agent-cap} are respectively set to 5 and 2 for both datasets ($5*2=10$). In other words, the first ten tweets would be randomly assigned to five agents. 

\subsection{Data streaming}
\label{datastream}
In this step, data are fetched from a data stream, e.g. Twitter. Such data sources have unique characteristics, in contrast to batch data sources, access to the dataset is limited by time--the whole data do not exist in advance. In other words, the system receives one data point at a time.
In this paper, two streaming data sources are used for topic detection: the COVID-19 dataset and the FA CUP dataset. These datasets have been explained in section \ref{dataset}. The streaming step fetches tweets from the data stream, and after passing a tweet through a preprocessing channel, assigns it to one of the agents (clusters).

\subsubsection{Data preprocessing}
\label{preprocessing}
Because of the short length of tweets, Twitter users usually prefer to use informal language. Due to this informality, tweets should be first preprocessed to be prepared for further processing. In this phase, several subtasks are applied to the newly-arrived tweet before passing it to the next step (Data assignment).
\begin{itemize}
\item URL removal: In this subtask, hyperlinks to various webpages are removed, as they generally do not contribute to the topic.
\item Hashtag tokenization: Hashtags are words or phrases starting with a '\#' character. Hashtags may contribute to the tweet's topic as they carry contextual information. A hashtag semantically links a tweet to all other tweets including it. As a hashtag usually consists of several words, in this subtask, it is separated into its constituting words.
\item Mention removal: Twitter users might be mentioned in a tweet with an '@' character followed by their username. These names are also removed from tweets as they usually do not contain useful information for topic detection. 
\item Tweet cleaning: As explained above, Informal language is often preferred in Twitter, which tends to include digits or special characters such as ``[]()/!?.''. Such characters are also removed in this subtask.
\item Tweet tokenization: Finally, the cleaned tweets are tokenized using the whitespace character. Bigrams and trigrams including a hyphen between words are left unchanged as they usually convey a non-compositional phrase, e.g., ``brother-in-law'', or ``chronicle-independent''. Moreover, all capital words are lowered.
\end{itemize}

\subsection{Data assignment}
\label{dataassignment}
In this step, the coordinator receives a tweet and assigns it to one of the existing clusters that is semantically the most similar to it. The coordinator is a component for monitoring the system and collaborating with agents (clusters). 
The similarity between a tweet and a cluster is measured by the similarity of topics covered by them. This similarity is being measured according to a state of the art sentence embedding model. More specifically, we first compute the sentence embedding vector of a given tweet based on the Language Agnostic Bert Sentence Embedding (LaBSE) model \cite{feng2020language}, which generates similar embeddings for bilingual or monolingual sentence pairs that are semantically similar. We then measure the cosine similarity of this vector with centroids of the existing clusters--the average sentence embedding vector of each cluster--to find the most similar cluster to the given tweet. The LaBSE \footnote{{https://tfhub.dev/google/LaBSE/1}} is a recently proposed multilingual sentence encoding model based on BERT \cite{devlin2018bert}, and its architecture is based on Bidirectional Dual-Encoder \cite{guo-etal-2018-effective} with Additive Margin Softmax \cite{ijcai2019-746}. It has been trained on 17 billion monolingual sentences and 6 billion bilingual sentence pairs. It is the state of the art model for measuring the semantic similarity between two documents/sentences, and can measure the semantic similarity of two sentences/documents even if they do not share any word. Moreover, this model takes the whole message conveyed by a tweet, instead of taking the constituting words of a tweet in isolation. 
Note that if the semantic distance of the newly arrived tweet from the most similar cluster to it is greater than a given threshold parameter, namely, \emph{assign radius}, a new cluster including only the new tweet will be generated. 
The cosine similarity measure in equation \ref{cosine} is used to compute the semantic similarity of two tweets, which are represented by two \emph{n}-dimensional vectors.
$\vec {tw}$ and $\vec {cl}$ respectively represent the tweet and cluster vectors.  

\label{cosine}
$cos(\vec {tw},\vec {cl}) =  \frac{\vec {tw} \cdot \vec {cl}}{{|\vec {tw}||\vec {cl}|}}$
$ = \frac{\vec {tw}}{|\vec {tw}|} \cdot \frac{\vec {cl}}{|\vec {cl}|}$
$= \frac{\sum_{i \in tw \cap cl} {tw}_i {cl}_i}{\sqrt{\sum_{i = 1}^{|tw|}} {tw}_{i}^2 \sqrt{\sum_{i = 1}^{|cl|} {cl}_{i}^2}}$



To prevent overflow in agents, we use a sliding window method. This window, indicated by a parameter named \emph{slid-win-init}, is set to twenty-four hours, and 1 minute, respectively in the COVID-19 and the FA CUP datasets' evaluations. This parameter differs from the \emph{timeslot} parameter which indicates a time interval, after which, the output of the system is stored and evaluated. In other words, topics and their keywords are separately extracted for each day in the COVID-19 and each minute in the FA CUP datasets. After each timeslot, a constant number of topics and a constant number of keywords per topic are stored. These constant numbers are represented by two parameters: \emph{no-topics} and \emph{no-keywords}. The \emph{no-topics} parameter varies in different experiments, whereas the \emph{no-keywords} is set to 5 and 9, respectively in the COVID-19 and the FA CUP datasets.

\subsection{Data exchange}
\label{dataexchange}
After assigning the tweets to existing clusters in the data assignment step, a multi-agent communication-based data exchange occurs periodically among agents under the supervision of the coordinator. This period is a parameter in our approach, named \emph{comm-int}. In this step, the coordinator redistributes outlier tweets among clusters to achieve higher cluster density. 
Concretely, in each timeslot, the agents determine their outlier tweets and return them back to the coordinator. A tweet is assumed an outlier in a cluster if its cosine similarity from the cluster centroid is lower than a given threshold. This threshold, named \emph{outlier-threshold} is another parameter, which is set to $0.78$ and $0.73$ respectively in the COVID-19 and the FA CUP datasets.
Then, the coordinator redistributes these outliers among existing clusters (agents), again based on the cosine similarity. The intuition behind this communication is that due to the automatic update in cluster centroids, which is caused by the newly-added tweets, some tweets inside clusters gradually become an outlier. In other words, the topic carried by an outlier gradually gets away from the overall topic of the cluster including it. 
At the end of each communication phase, the weight of each agent is reduced by a parameter named \emph{agent-fading-rate}. After this update, if the weight of any agent is lower than a threshold, \emph{del-agent-weight}, it will be faded. This weight is incremented by each data point's arrival to an agent, but not decremented by each outlier's removal from that agent. 

As mentioned already, the proposed methodology includes several parameters, which have been initialized by the try-and-test method. The complete list of parameters, as well as their values separately for two datasets, are listed in Table \ref{table_parameters}. The proposed approach has two versions: ComStreamClust1 and ComStreamClust2.  The difference between these two versions is in parameters assign-radius and outlier-threshold. The value of assign-radius and outlier-threshold are 0.25 and 0.27 in ComStreamClust1 and 0.27 and 0.29 in ComStreamClust2 respectively.  
\begin{table*}[t]
\caption{The parameters used in the proposed methodology. cc, dp, ts, and kw respectively stand for cluster-center, data-point, time-slot and keyword. Values inside parantheses have been used in two versions of the proposed approach.}
\centering
\begin{tabular}{l l l l}
\hline
\bf Parameter name &\bf COVID-19&\bf FA CUP&\bf explanation \\ [0.5ex]
\hline
init-agents& 5 & 2 & initial number of agents\\
init-agent-cap& 5 & 2 & initial \# of dps per agents\\
timeslot & 24h& 1m & time-interval to store the output\\
comm-int & 1.5h& 1m & time-interval to repeat comm. phase\\
slid-win-int & 24h& 1m & time-interval for naming a dp is as old\\
assign-radius& 0.2 & (0.25,0.27) & max distance for assigning a dp to an agent\\
outlier-threshold& 0.22 & (0.27,0.29) & min dist. from cc for a dp to be an outlier\\
no-topics & 2-20& 2-20 &\# of topics stored in each ts\\
no-keywords& 5 & 9 &\# of kws per topic stored in each ts\\
agent-fading-rate& 0.5 & 0 &percentile of agents faded in comm. phase\\
del-agent-weight-threshold& 0.4 & 0 &weight threshold for deleting agents\\
 \hline
\end{tabular}
\label{table_parameters}
\end{table*}

\section{Experimental Evaluation}
\label{expeval}
In this section, we evaluate the proposed approach on two datasets, the COVID-19 and the FA CUP. Evaluation metrics, ground-truth, datasets, and obtained results have been explained with details in the following subsections.

{\bf Evaluation metrics:} We use topic recall, keyword recall, and keyword precision for evaluation. F-score is also used for keyword evaluation as the harmonic mean of precision and recall. These metrics are calculated according to equations \ref{prec} through \ref{fscore}. Keyword precision is the ratio of correctly extracted keywords of a topic over the total number of extracted keywords for that topic. Topic recall is the ratio of correctly extracted topics over the total number of topics. 
Keyword recall is the ratio of correctly extracted keywords of a topic over the total number of keywords of that topic. The correct topics and their keywords have been collected in a ground-truth set. Note that we did not use topic precision, because there exist hot topics in the datasets which might not be in the ground-truth set. For example, daily events such as the death of someone's cat might be extracted as a hot topic, where it is not relevant to any topic in the ground-truth.
\begin{equation}
\label{prec}
\displaystyle  
P_{kw} = \frac{\textit{ \# of estimated kws for $T_i$}}{\textit{\# of kws for $T_i$ in ground-truth}}
\end{equation}

\begin{equation}
\label{recall}
\displaystyle  R_{tp} = \frac{\textit{\# of correctly extracted topics}}{\textit{\# of topics in ground-truth}}\\
\end{equation}
\begin{equation}
\label{recallkw}
\displaystyle R_{kw} = \frac{\textit{\# of correctly extracted kws for $T_i$}}{\textit{\# of kws for $T_i$ in ground-truth}}
\end{equation}

\begin{equation}
\label{fscore}
\displaystyle  Fscore_{P,R} = \frac{2*P*R}{P+R}
\end{equation}

\textbf{Ground-truth:} The ground-truth data are available for the FA CUP dataset, which have been explained with details in \cite{aiello2013sensing}. However, to the best knowledge of the authors, there is no ground-truth publicly available for the COVID-19 dataset. To generate a ground-truth for this dataset, we manually extracted the hot topics and their associated keywords from online media and search engines, separately for each day from March 29 to April 30. We have made these ground-truth data publicly available  \footnote{\url{https://www.kaggle.com/thelonecoder/labelled-1000k-covid19-dataset}}.

\begin{figure*}[t]
\includegraphics[width=\textwidth]{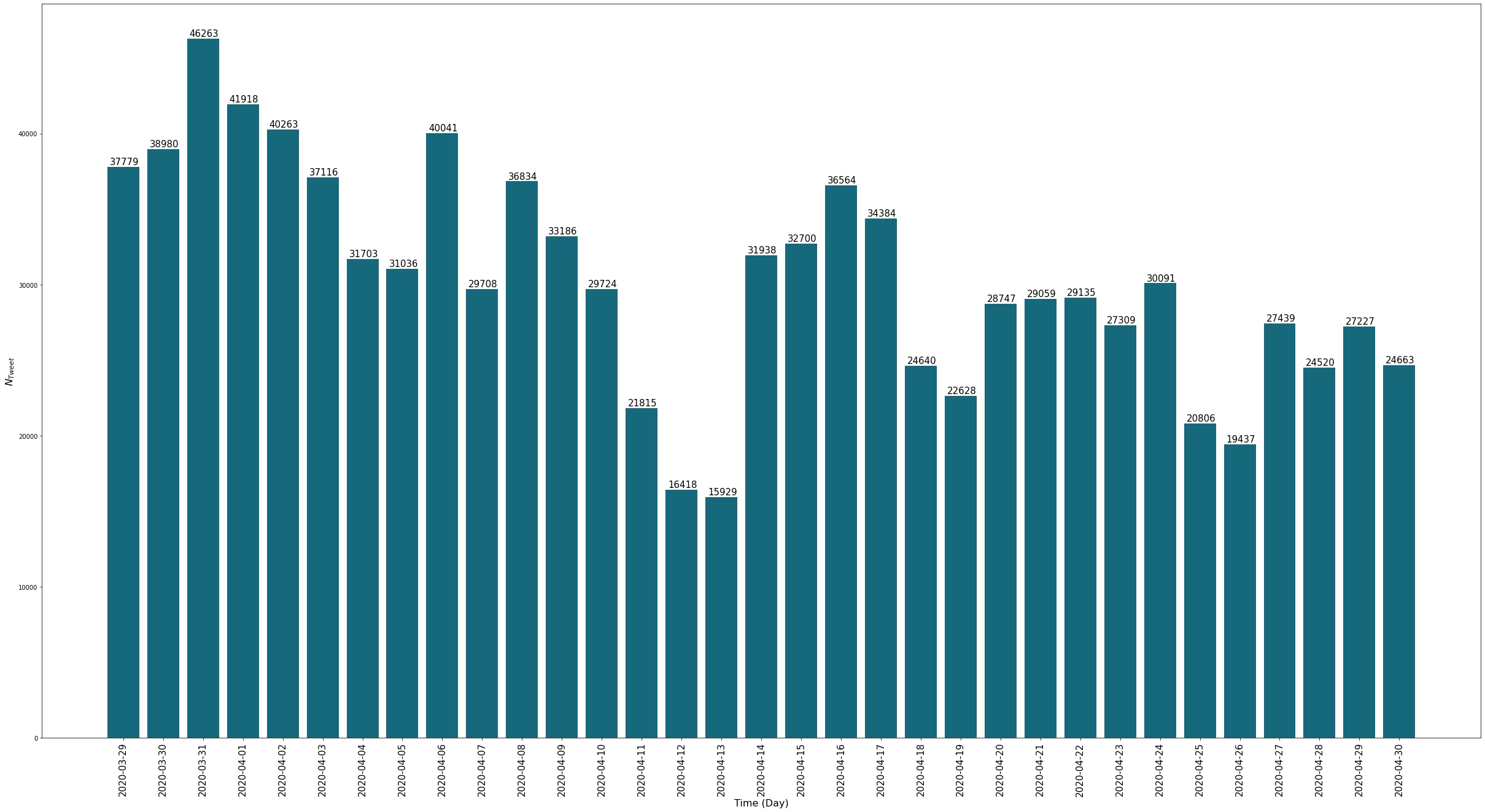}
\captionsetup{justification=centering}
\centering
\caption{Daily distribution of COVID-19 tweets in a 31-day interval.}
\label{NTweetCOV}
\end{figure*}    

\begin{figure*}[t]
\centering
\includegraphics[width=\textwidth ]{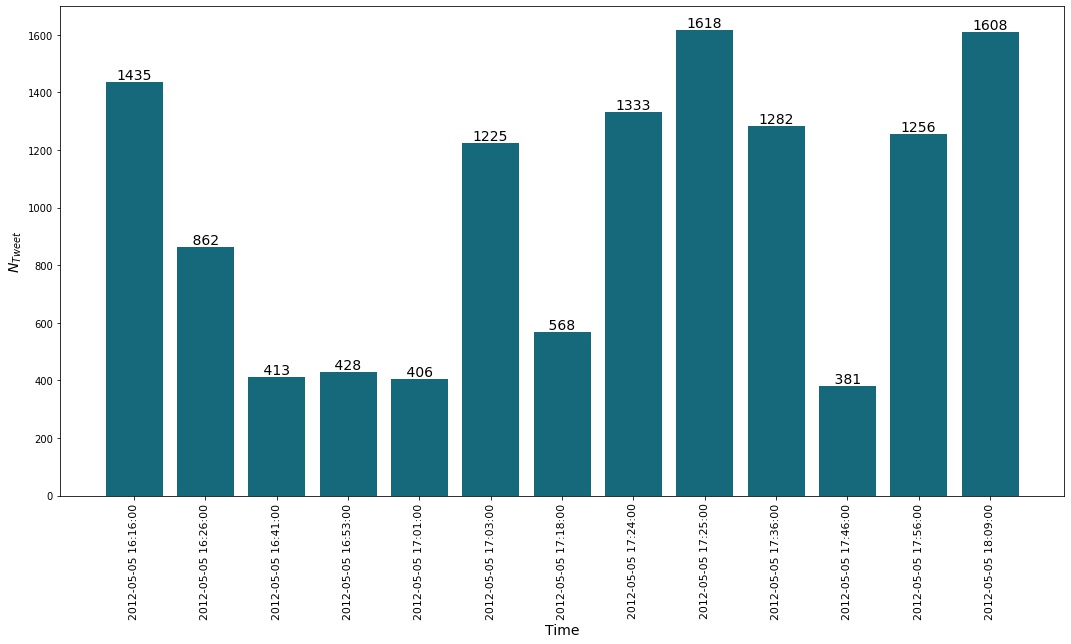}
\captionsetup{justification=centering}
\caption{Daily distribution of the FA CUP tweets in a 13 timeslots including special events.}
\label{NTweetFACUP}
\end{figure*}    

\subsection{Dataset}
\label{dataset}
As already mentioned, two datasets have been used in this paper. The Football Association Challenge Cup, FA CUP, was compiled during a football game between Chelsea and Liverpool, on May 05, 2012, from 16:00:00 to 18:30:00. The data have been crawled using key hashtags such as the event name and the name of teams and key players. The set of hot topics in this dataset is comprised of 13 special times including the start and the end of the match, the goal times, penalizing the players, and so on. This dataset includes about 113 thousand English tweets already labeled in \cite{aiello2013sensing}. 

The COVID-19 dataset is a collection of tweets compiled from Twitter from March 29 to April 30, 2020. This dataset has been assembled by crawling tweets including the hashtag \#COVID19.
The total number of tweets in this dataset is about 9 million but we randomly chose 1 million tweets and manually identified thirty topics in them as our groundtruth. We have made this subset publicly available for other researchers' use 
\footnote{\url{https://www.kaggle.com/thelonecoder/labelled-1000k-covid19-dataset}}
The number of tweets in each timeslot for both datasets has been illustrated as a histogram in Figures \ref{NTweetCOV} and \ref{NTweetFACUP}. Both datasets have been automatically collected by using hashtags, therefore irrelevant (outlier) tweets might exist in them which would lead to challenges in topic detection. This issue was the reason for neglecting topic precision as an evaluation metric.

\subsection{Results}
\begin{figure*}[t]
\includegraphics[width=\textwidth]{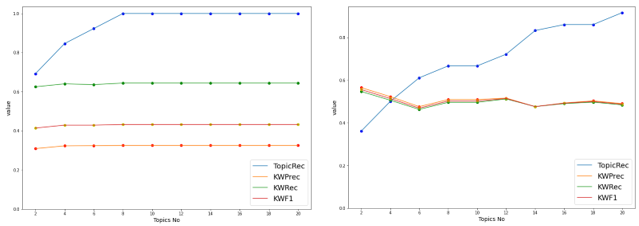}
\captionsetup{justification=centering}
\caption{TRec, KWRec and KWPrec for different number of topics per timeslot in the FA CUP (left) and COVID-19 (right) dataset.}
\label{TopicNo}
\end{figure*}

The evaluation metrics include topic and keyword recall, keyword precision, and F-score (for keyword evaluation). 
We used micro-averaging for computing the final value of evaluation metrics both for the topic and keyword evaluation. 
We conducted several experiments, the results of which are portrayed in Figure \ref{TopicNo}. These results have been obtained when {topic-number-per-timeslot} is 2 to 20 (for both datasets) and {keyword-number-per-topic} is 5 for the covid-19 and 9 for the FA CUP datasets. The exact values of topic recalls have been also provided in Table \ref{table_results}. 

It can be concluded from this table and figure that a higher number of topics would result in a higher topic recall, especially in the COVID-19 dataset. The intuition behind the harsh slope of the line (from 0 to 3) in the COVID-19, is that the number of hot topics per timeslot ranges from 0 to 3, and therefore increasing the number of estimated hot topics would increase topic recall until 3, but after 3, due to the lower number of topics per timeslot in the ground-truth, the line rises with a lower slope. However, increasing this number does not affect other metrics (Keyword precision and recall) much, because having more topics would require estimating more topic keywords, whereas our estimation would not be always correct. 
\begin{table*}
\caption{Obtained results for  both datasets by the proposed approach.}
\begin{tabular}{ l l l l l l l l l l}
\hline
2&4&6&8&10&12&14&16&18&20\\
\hline
&&&&COVID-19&&&&&\\
{0.361}&{0.500}&{0.611}&0.667&0.667&0.722&{0.833}&{0.861}&{0.861}&{0.917}\\
\hline
&&&&FA CUP&&&&&\\
0.692&0.840&0.923&1&1&1&1&1&1&1\\
\hline
\end{tabular}
\label{table_results}
\end{table*}

\begin{figure*}[t]
\centering
\includegraphics[width=\textwidth]{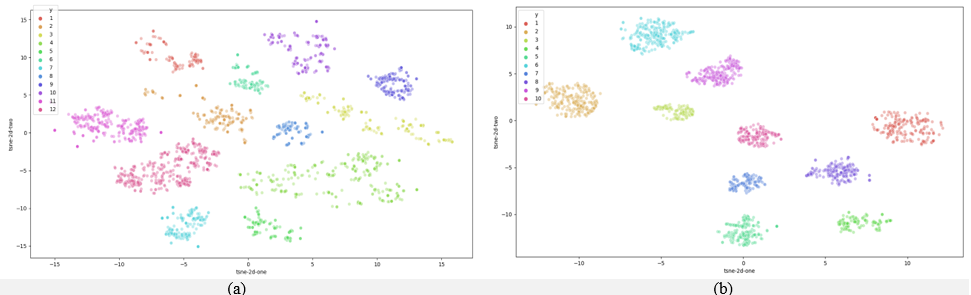}
\captionsetup{justification=centering}
\caption{The t-distributed stochastic neighbor embedding (tSNE) diagrams for a specific timeslot in the FA CUP (a) and the COVID-19 (b) datasets.}
\label{TSNE}
\end{figure*}

Finally, we provide the t-distributed stochastic neighbor embedding (tSNE) diagrams for both datasets in a specific timeslot in Figure \ref{TSNE}. Each agent in this diagram represents a cluster including similar data-points, i.e., tweets sharing a topic. 

\subsection{Discussion and comparison}
\label{discussin}
We obtained different topic recall values for different values for topic-number-per-timeslot, which have been provided in Table \ref{table_comparison}. This table also compares the obtained values with other approaches applied to the FA CUP dataset for topic detection. The main reason for applying the proposed approach on the FA CUP dataset is fair comparison because the same set of data-points are used in the FACUP by different researchers; However, in the case of COVID-19, a different dataset has been used in each research work. Moreover, people usually do not make their dataset publicly available. As mentioned already, the difference between two versions of ComStreamClust is in two parameters, assign-radius and outlier-threshold. ComStreamClust1 with lower values for these parameters starts not very good but obtains the best result fast. On the other hand, ComStreamClust2 with higher values for these parameters starts good but its improvement speed is lower. The intuition behind this issue is that the lower radius causes higher number of agents, and higher number of agents would make them more specific. But when the radius is higher, lower number of agents with more generality would be generated. 

Several erroneous cases were causing lower topic recall in both datasets, such as the overshadowing phenomenon. In the FA CUP, for instance, a goal was achieved by Drogba in the 24th minute which is a hot topic in timeslot 24. There is another hot topic in the 25th minute that discusses passing the ball to Drogba before achieving the goal. The latter topic was overshadowed by the first one, i.e., the latter was lost in minute 25 among tweets issued in minute 24. 

\begin{table*}[t]
\caption{The obtained results of the proposed approach and its comparison with other methods when applying on the FA CUP dataset with different topic numbers. TR stands for Topic Recall. The maximum value in each column is in bold.}
\centering
\begin{tabular}{l l l l l l l l l l l}
\hline
Method&TR@2&TR@4&TR@6&TR@8&TR@10&TR@12&TR@14&TR@16&TR@18&TR@20\\
\hline
&&&&&FA CUP&&&&\\
Gfeat-P&0.000&0.308&0.308&0.375&0.375&0.375&0.375&0.375&0.375&0.375\\
LDA&0.692&0.692&0.840&0.840&0.923&0.923&0.840&0.840&0.840&0.750\\
BNgram&0.769&{\bf 0.923}&{\bf 0.923}&0.923&0.923&0.923&0.923&0.923&0.923&0.923\\
SFPM&0.615&0.840&0.840&{\bf 1}&{\bf 1}&{\bf 1}&{\bf 1}&{\bf 1}&{\bf 1}&{\bf 1}\\
Doc-P&0.769&0.840&{\bf 0.923}&0.923&{\bf 1}&{\bf 1}&{\bf 1}&{\bf 1}&{\bf 1}&{\bf 1}\\
ComStreamClust1&{0.692}&{ 0.840}&{\bf 0.923}&{\bf1}&{\bf1}&{\bf1}&{\bf 1}&{\bf 1}&{\bf 1}&{\bf 1}\\
ComStreamClust2&{\bf 0.840}&{\bf 0.923}&{\bf 0.923}&0.923&0.923&{0.923}&{\bf 1}&{\bf 1}&{\bf 1}&{\bf 1}\\

\hline
\end{tabular}
\label{table_comparison}
\end{table*}

\begin{figure*}[t]
\centering
\includegraphics[height=3in]{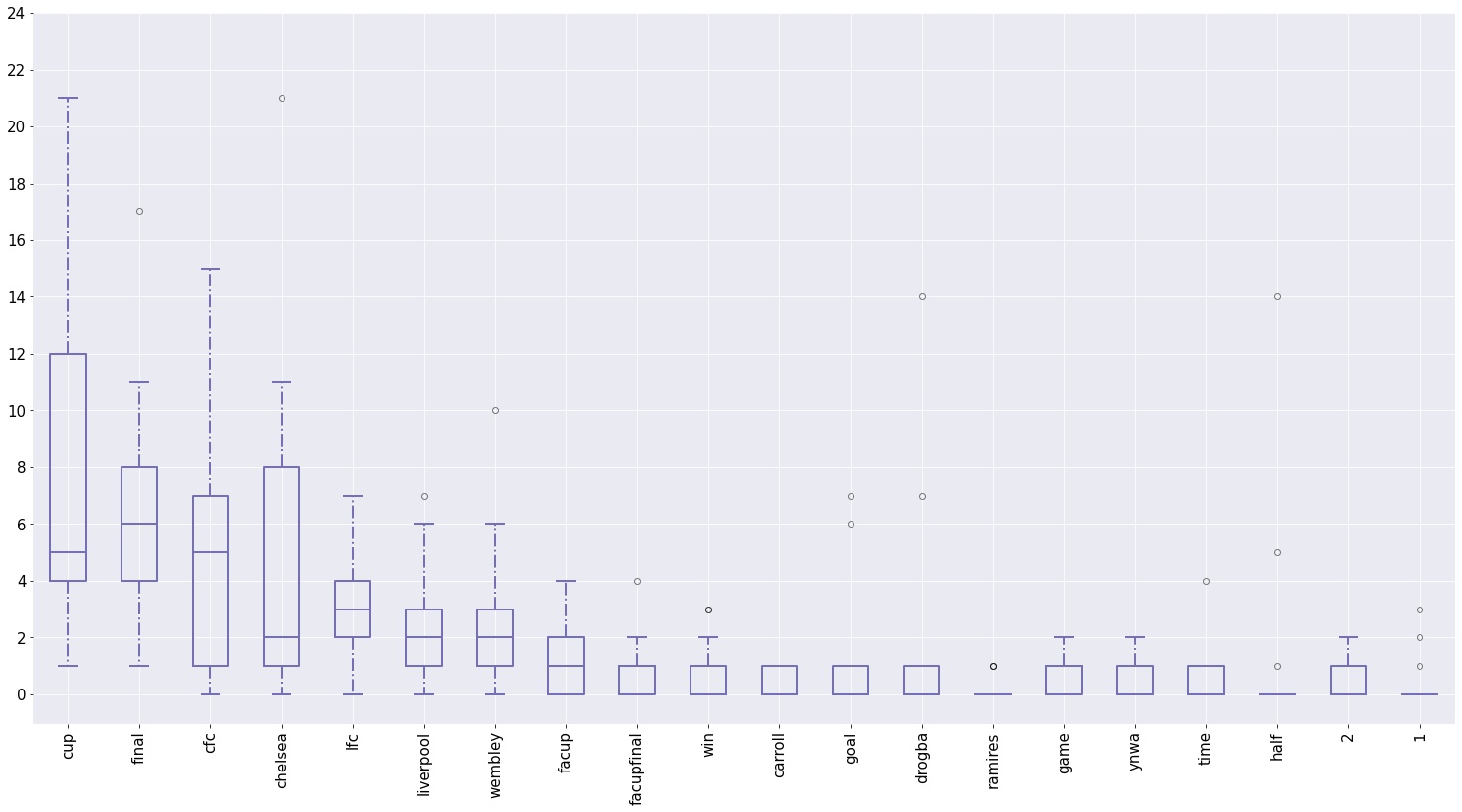}
\captionsetup{justification=centering}
\caption{The Boxplot of keywords in the COVID-19 dataset.}
\label{boxkwfacup}
\end{figure*}

\begin{figure*}[t]
\centering
\includegraphics[width=5.5in]{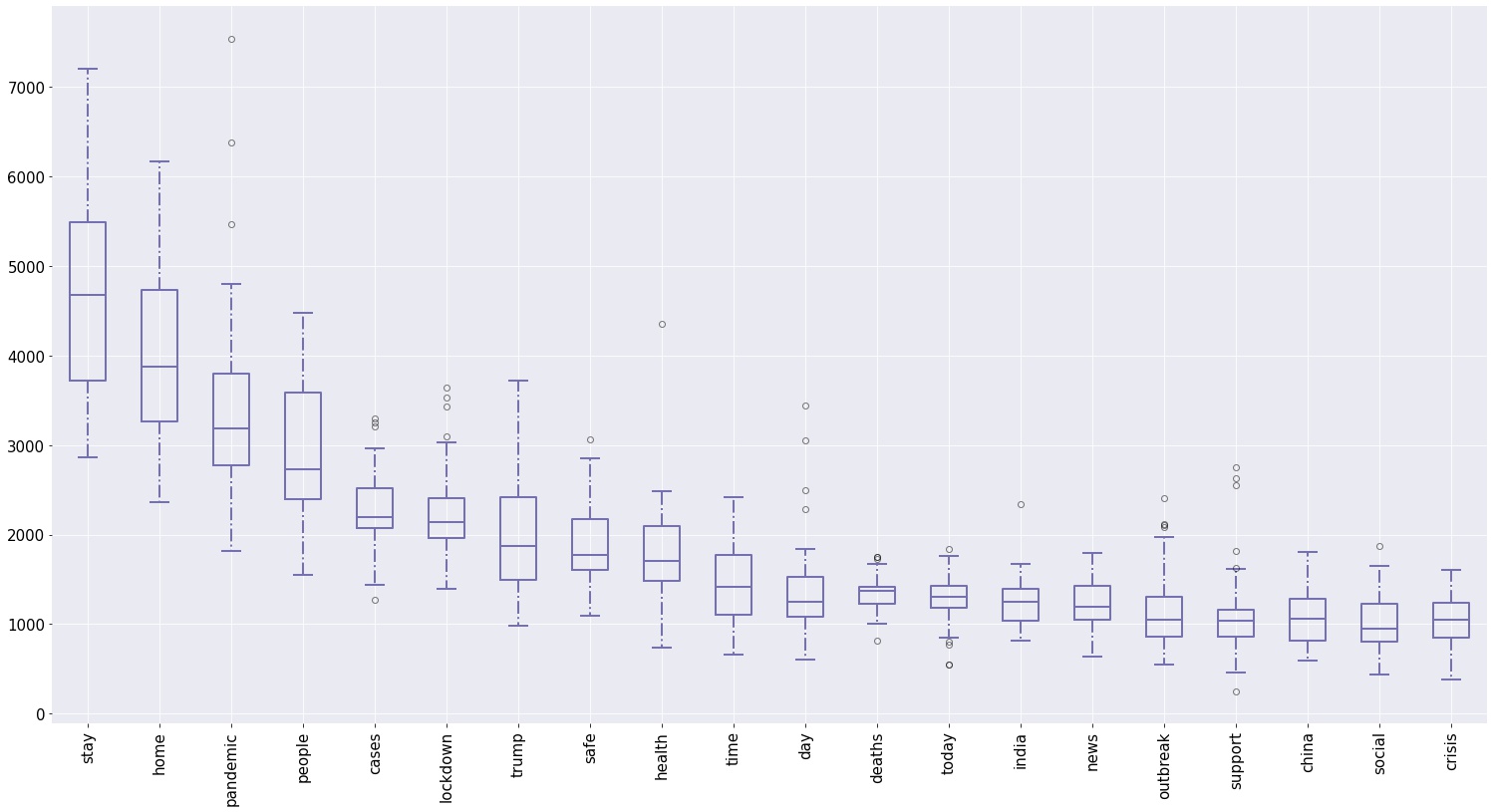}
\captionsetup{justification=centering}
\caption{The Boxplot of keywords in the COVID-19 dataset.}
\label{boxkwCOVID19}
\end{figure*}

{\bf Keyword analysis:} We tested the proposed approach with different values for the parameter, \emph{no-keywords} (per topic), and concluded that 5 and 9  keywords per topic respectively in the FACUP and COVID-19 datasets achieve the best results. A lower number of keywords would result in lower topic-recall due to detecting fewer topics, but higher topic precision. A greater number of keywords, on the other hand, would detect more topics causing higher topic recall and lower topic precision. At the end of all time-slots in each dataset, some keywords have been labeled as most frequent, which have been illustrated as boxplots in Figures \ref{boxkwfacup} and \ref{boxkwCOVID19}.

As can be seen in COVID-19's boxplot, the most frequent words are ``stay'', ``pandemic'', ``home'' and ``people'', which may imply that due to the COVID-19 pandemic, people suggest each other to ``stay at home''. We observed in this dataset that the majority of tweets include the hashtag \#StayAtHome. Note that obvious frequent keywords such as ``COVID19'', ``coronavirus'', ``corona'', and ``virus'' have been treated as stopwords, as they appear in almost all tweets.   
We chose a few sample tweets from each dataset, which include a hot topic in its timeslot. Table \ref{table_samples} lists these sample tweets.

\begin{table*}[t]
\caption{Sample tweets including hot topics for each dataset.}
\centering
\begin{tabular}{l l l}
\hline
Text&Date-Time&keywords\\ 
\hline
COVID-19&&\\
\hline
\#BREAKING The \#UK \#PrimeMinister @BorisJohnson has been&19:25:28&icu hospital\\
moved to \#ICU "\#BorisJohnson moved to intensive care after being &2020-04-06&Boris johnson\\admitted to hospital with \#coronavirus symptoms" \#COVID19 \\https://t.co/mq4gDedDGx\\
\\
It’s \#EarthDay2020. It’s ironic that the \#COVID19 crisis has made us think&03:18:01&earth change\\
more about how vulnerable we are on this fragile earth. Maybe now our&2020-04-22&climate\\
governments will rethink their intransigence on climate change action.\\
\hline
FA CUP&&\\
\hline
44' Daniel Agger made a hard tackle to Mikel. And shown yellow card&17:01:34&mikel daniel\\ by the referee. \#FACup&May 05 2012&agger yellow card\\
\\
\o/ Yay Chelsea! RT @itvfootball: Congrats to \#CFC on beating \#LFC 2-1&18:09:52&cup chelsea\\ 
and winning the 2012 \#FACupFinal&May 05 2012&champions\\
\hline
\end{tabular}
\label{table_samples}
\end{table*}

{\bf Multi-agent setup:} The proposed approach is capable of being parallelized. In order to prove this claim, we implemented it in the elixir language \footnote{\url{https://elixir-lang.org/docs.html}} using a multi-processor system, after running the implemented system in python on a single-processor system. The cpu Specifications for this system is 4x i7 6700 HQ @3.8GHz and its RAM is 16 GB. We conducted this experiment to see if the proposed approach can be executed in a multi-agent system and also how much time can be saved in a parallel setting compared to the sequential case. More specifically, we used a multi-agent system including eight processors, each of which handles one cluster. If the number of clusters is more than eight (which is the case in some time intervals), the parallel system would transform to an eight-processor pipeline --the ninth and other clusters will be concurrently processed.
The time intervals spent for the sequential and parallel cases are respectively 5 minutes and 14 seconds and 4 minutes and 40 seconds.  Note that sequential processing in python takes the advantage of the optimal implementation of libraries such as numpy which drastically decreases the execution time, whereas elixir lacks such optimality. Every process in this language has to be executed in parallel; no sequential execution is allowed in it. In conclusion, due to the facts discussed above, the improvement in the execution time is not substantial. This improvement could be higher with higher-speed processors.

\begin{table*}[t]
\caption{Sample tweets turned to normal data-points from outlier.}
\centering
\begin{tabular}{l l l}
\hline
Tweet&Tweet assumed as outlier&Tweet assumed as normal\\ 
\hline
COVID-19&&\\
\hline
british prime minister boris &['boris', 'johnson', 'stay', 'pandemic', 'trump',&['boris', 'johnson', 'care', 'intensive', 'stay',
\\
johnson moved intensive care&'news', 'care', 'social', 'lockdown', 'health']& 'moved', 'prime', 'minister', 'icu', 'recovery']\\ 
\hline
FACUP&&\\
great save by cech from a low &['the', 'arsenal', 'you', 'goal', 'cfc',&['cup', 'fa', 'final', 'wembley', 'cfc', \\
suarez shot. fa cup final&'still', 'please', 'minutes', 'have', 'chelsea']& 'sl', 'chelsea', 'cech', 'suarez', 'save']\\
\hline
\end{tabular}
\label{table_outliernormal}
\end{table*}
{\bf Data-point tracking:} For deeper analysis, we tracked some data-points during their life-cycle in our methodology. Some data-points might be assumed as an outlier in a cluster but after reassignment to another cluster, they turn to be a normal data-point. Two sample tweets have been tracked in Table \ref{table_outliernormal}. In this table, the COVID-19 tweet was first categorized in a general-concept cluster, but when it was identified as an outlier in that cluster, the coordinator re-assigned it to a more specific cluster; Therefore, a normal data-point could be kept among clustered data. In the FACUP tweet, the situation is also similar. Saving the shot of Suarez by Cech was supposed to be an outlier in a general tweet and then labeled as a normal data-point in a more specific one. 

The strengths of the proposed approach include its dynamic, communicative and parllelizable nature and keeping the detected topics (clusters) as pure as possible. ComStreamClust attempts to keep clusters fresh, by discarding older tweets, and updating the clusters by adding newer ones, and also detecting and deleting the outliers. Note that a tweet might gradually turn into an outlier, due to the updates that happen to the cluster including it. The weaknesses of this approach might be its need for parameter tuning, i.e., it requires adapting each parameter for the given dataset.

\section{Conclusion and future work}
This paper proposes a new topic detection approach using stream clustering on Twitter data. The proposed approach, named ``ComStreamClust'', is unique in that it benefits from a communication phase, in which, clusters communicate with each other in a multi-agent and parallelizable setting. ComStreamClust has been applied on two datasets, the COVID-19 and the FA CUP. When applied on the FA CUP dataset, it was shown that the proposed methodology provides superior or in some cases, equal perdormance compared to other methodologies. The current analysis on the COVID-19 dataset approves the assumption that social media can help governments and health centers cure this pandemic in a more efficient and rapid manner. For example, almost all Twitter users have used \#StayAtHome in their tweets, which in turn would remind people that staying at home is the most efficient treatment for the COVID-19 pandemic. Our future works include exploiting images inside tweets to accomplish a multinodal topic detection on Twitter.

\bibliographystyle{spbasic}

\bibliography{refs}

\begin{thebibliography}{29}
\providecommand{\natexlab}[1]{#1}
\providecommand{\url}[1]{{#1}}
\providecommand{\urlprefix}{URL }
\expandafter\ifx\csname urlstyle\endcsname\relax
  \providecommand{\doi}[1]{DOI~\discretionary{}{}{}#1}\else
  \providecommand{\doi}{DOI~\discretionary{}{}{}\begingroup
  \urlstyle{rm}\Url}\fi
\providecommand{\eprint}[2][]{\url{#2}}

\bibitem[{Aiello et~al.(2013)Aiello, Petkos, Martin, Corney, Papadopoulos,
  Skraba, G{\"o}ker, Kompatsiaris, and Jaimes}]{aiello2013sensing}
Aiello LM, Petkos G, Martin C, Corney D, Papadopoulos S, Skraba R, G{\"o}ker A,
  Kompatsiaris I, Jaimes A (2013) Sensing trending topics in twitter. IEEE
  Transactions on Multimedia 15(6):1268--1282

\bibitem[{Asgari-Chenaghlu et~al.(2020)Asgari-Chenaghlu, Nikzad-Khasmakhi, and
  Minaee}]{asgari2020COVID}
Asgari-Chenaghlu M, Nikzad-Khasmakhi N, Minaee S (2020) Covid-transformer:
  Detecting trending topics on twitter using universal sentence encoder. arXiv
  preprint arXiv:200903947

\bibitem[{Atefeh and Khreich(2015)}]{atefeh2015survey}
Atefeh F, Khreich W (2015) A survey of techniques for event detection in
  twitter. Computational Intelligence 31(1):132--164

\bibitem[{Blei et~al.(2003)Blei, Ng, and Jordan}]{blei2003latent}
Blei DM, Ng AY, Jordan MI (2003) Latent dirichlet allocation. Journal of
  machine Learning research 3(Jan):993--1022

\bibitem[{Carnein et~al.(2017)Carnein, Assenmacher, and
  Trautmann}]{carnein2017stream}
Carnein M, Assenmacher D, Trautmann H (2017) Stream clustering of chat messages
  with applications to twitch streams. In: International Conference on
  Conceptual Modeling, Springer, pp 79--88

\bibitem[{Cataldi et~al.(2010)Cataldi, Di~Caro, and
  Schifanella}]{cataldi2010emerging}
Cataldi M, Di~Caro L, Schifanella C (2010) Emerging topic detection on twitter
  based on temporal and social terms evaluation. In: Proceedings of the tenth
  international workshop on multimedia data mining, pp 1--10

\bibitem[{Cer et~al.(2018)Cer, Yang, Kong, Hua, Limtiaco, John, Constant,
  Guajardo-Cespedes, Yuan, Tar et~al.}]{cer2018universal}
Cer D, Yang Y, Kong Sy, Hua N, Limtiaco N, John RS, Constant N,
  Guajardo-Cespedes M, Yuan S, Tar C, et~al. (2018) Universal sentence encoder.
  arXiv preprint arXiv:180311175

\bibitem[{Cigarr{\'a}n et~al.(2016)Cigarr{\'a}n, Castellanos, and
  Garc{\'\i}a-Serrano}]{cigarran2016step}
Cigarr{\'a}n J, Castellanos {\'A}, Garc{\'\i}a-Serrano A (2016) A step forward
  for topic detection in twitter: An fca-based approach. Expert Systems with
  Applications 57:21--36

\bibitem[{Dehkharghani and Yilmaz(2013)}]{dehkharghani2013automatically}
Dehkharghani R, Yilmaz C (2013) Automatically identifying a software product's
  quality attributes through sentiment analysis of tweets. In: 2013 1st
  International Workshop on Natural Language Analysis in Software Engineering
  (NaturaLiSE), IEEE, pp 25--30

\bibitem[{Dehkharghani et~al.(2014)Dehkharghani, Mercan, Javeed, and
  Saygin}]{dehkharghani2014sentimental}
Dehkharghani R, Mercan H, Javeed A, Saygin Y (2014) Sentimental causal rule
  discovery from twitter. Expert Systems with Applications 41(10):4950--4958

\bibitem[{Devlin et~al.(2018)Devlin, Chang, Lee, and
  Toutanova}]{devlin2018bert}
Devlin J, Chang MW, Lee K, Toutanova K (2018) Bert: Pre-training of deep
  bidirectional transformers for language understanding. arXiv preprint
  arXiv:181004805

\bibitem[{Fang et~al.(2014)Fang, Zhang, Ye, and Li}]{fang2014detecting}
Fang Y, Zhang H, Ye Y, Li X (2014) Detecting hot topics from twitter: A
  multiview approach. Journal of Information Science 40(5):578--593

\bibitem[{Feng et~al.(2020)Feng, Yang, Cer, Arivazhagan, and
  Wang}]{feng2020language}
Feng F, Yang Y, Cer D, Arivazhagan N, Wang W (2020) Language-agnostic bert
  sentence embedding. arXiv preprint arXiv:200701852

\bibitem[{Guo et~al.(2018)Guo, Shen, Yang, Ge, Cer, Hernandez~Abrego, Stevens,
  Constant, Sung, Strope, and Kurzweil}]{guo-etal-2018-effective}
Guo M, Shen Q, Yang Y, Ge H, Cer D, Hernandez~Abrego G, Stevens K, Constant N,
  Sung YH, Strope B, Kurzweil R (2018) Effective parallel corpus mining using
  bilingual sentence embeddings. In: Proceedings of the Third Conference on
  Machine Translation: Research Papers, Association for Computational
  Linguistics, Brussels, Belgium, pp 165--176, \doi{10.18653/v1/W18-6317},
  \urlprefix\url{https://www.aclweb.org/anthology/W18-6317}

\bibitem[{Guo et~al.(2013)Guo, Xiang, Chen, Huang, and Hao}]{guo2013lda}
Guo X, Xiang Y, Chen Q, Huang Z, Hao Y (2013) Lda-based online topic detection
  using tensor factorization. Journal of Information Science 39(4):459--469

\bibitem[{Hahsler and Bola{\~n}os(2016)}]{hahsler2016clustering}
Hahsler M, Bola{\~n}os M (2016) Clustering data streams based on shared density
  between micro-clusters. IEEE Transactions on Knowledge and Data Engineering
  28(6):1449--1461

\bibitem[{Hasan et~al.(2019)Hasan, Orgun, and Schwitter}]{hasan2019real}
Hasan M, Orgun MA, Schwitter R (2019) Real-time event detection from the
  twitter data stream using the twitternews+ framework. Information Processing
  \& Management 56(3):1146--1165

\bibitem[{Ibrahim et~al.(2018)Ibrahim, Elbagoury, Kamel, and
  Karray}]{ibrahim2018tools}
Ibrahim R, Elbagoury A, Kamel MS, Karray F (2018) Tools and approaches for
  topic detection from twitter streams: survey. Knowledge and Information
  Systems 54(3):511--539

\bibitem[{Mottaghinia et~al.(2020)Mottaghinia, Feizi-Derakhshi, Farzinvash, and
  Salehpour}]{mottaghinia2020review}
Mottaghinia Z, Feizi-Derakhshi MR, Farzinvash L, Salehpour P (2020) A review of
  approaches for topic detection in twitter. Journal of Experimental \&
  Theoretical Artificial Intelligence pp 1--27

\bibitem[{Nur'Aini et~al.(2015)Nur'Aini, Najahaty, Hidayati, Murfi, and
  Nurrohmah}]{nur2015combination}
Nur'Aini K, Najahaty I, Hidayati L, Murfi H, Nurrohmah S (2015) Combination of
  singular value decomposition and k-means clustering methods for topic
  detection on twitter. In: 2015 International Conference on Advanced Computer
  Science and Information Systems (ICACSIS), IEEE, pp 123--128

\bibitem[{Papka et~al.(1998)Papka, Allan et~al.}]{papka1998line}
Papka R, Allan J, et~al. (1998) On-line new event detection using single pass
  clustering. University of Massachusetts, Amherst 10(290941.290954)

\bibitem[{Petkos et~al.(2014)Petkos, Papadopoulos, and
  Kompatsiaris}]{petkos2014two}
Petkos G, Papadopoulos S, Kompatsiaris Y (2014) Two-level message clustering
  for topic detection in twitter. In: SNOW-DC@ WWW, pp 49--56

\bibitem[{Popescu and Pennacchiotti(2010)}]{popescu2010detecting}
Popescu AM, Pennacchiotti M (2010) Detecting controversial events from twitter.
  In: Proceedings of the 19th ACM international conference on Information and
  knowledge management, pp 1873--1876

\bibitem[{Rill et~al.(2014)Rill, Reinel, Scheidt, and Zicari}]{rill2014politwi}
Rill S, Reinel D, Scheidt J, Zicari RV (2014) Politwi: Early detection of
  emerging political topics on twitter and the impact on concept-level
  sentiment analysis. Knowledge-Based Systems 69:24--33

\bibitem[{Tembhurnikar and Patil(2015)}]{tembhurnikar2015topic}
Tembhurnikar SD, Patil NN (2015) Topic detection using bngram method and
  sentiment analysis on twitter dataset. In: 2015 4th International Conference
  on Reliability, Infocom Technologies and Optimization (ICRITO)(Trends and
  Future Directions), IEEE, pp 1--6

\bibitem[{Vaswani et~al.(2017)Vaswani, Shazeer, Parmar, Uszkoreit, Jones,
  Gomez, Kaiser, and Polosukhin}]{vaswani2017attention}
Vaswani A, Shazeer N, Parmar N, Uszkoreit J, Jones L, Gomez AN, Kaiser {\L},
  Polosukhin I (2017) Attention is all you need. In: Advances in neural
  information processing systems, pp 5998--6008

\bibitem[{Xie et~al.(2016)Xie, Zhu, Jiang, Lim, and Wang}]{xie2016topicsketch}
Xie W, Zhu F, Jiang J, Lim EP, Wang K (2016) Topicsketch: Real-time bursty
  topic detection from twitter. IEEE Transactions on Knowledge and Data
  Engineering 28(8):2216--2229

\bibitem[{Yang et~al.(2019)Yang, Hernandez~Abrego, Yuan, Guo, Shen, Cer, Sung,
  Strope, and Kurzweil}]{ijcai2019-746}
Yang Y, Hernandez~Abrego G, Yuan S, Guo M, Shen Q, Cer D, Sung Yh, Strope B,
  Kurzweil R (2019) Improving multilingual sentence embedding using
  bi-directional dual encoder with additive margin softmax. In: Proceedings of
  the Twenty-Eighth International Joint Conference on Artificial Intelligence,
  {IJCAI-19}, International Joint Conferences on Artificial Intelligence
  Organization, pp 5370--5378, \doi{10.24963/ijcai.2019/746},
  \urlprefix\url{https://doi.org/10.24963/ijcai.2019/746}

\bibitem[{Zhou et~al.(2011)Zhou, Zhong, and Li}]{zhou2011hot}
Zhou E, Zhong N, Li Y (2011) Hot topic detection in professional blogs. In:
  International Conference on Active Media Technology, Springer, pp 141--152

\end{thebibliography}
\end{document}